\documentclass[aps,prd,twocolumn,groupedaddress,eqsecnum,amssymb,showpacs,
nofootinbib]{revtex4}
\usepackage{graphicx,mathptm,bm,times,epsfig}

\newcommand{\be}{\begin{equation}}
\newcommand{\ee}{\end{equation}}
\newcommand{\bea}{\begin{eqnarray}}
\newcommand{\eea}{\end{eqnarray}}
\newcommand{\rhostar}{{ \rho_\ast }} 
\newcommand{\rhowiggle}{{ {\widetilde \rho} }} 

\begin{document}

\title{Limitations of anthropic predictions for the cosmological
constant $\Lambda$:\\ Cosmic Heat Death of Anthropic Observers}

\author{Laura Mersini-Houghton}
\email[]{mersini@physics.unc.edu}
\affiliation{Department of Physics and Astronomy, UNC-Chapel Hill,
NC, 27599-3255, USA}

\author{Fred C. Adams} 
\email[]{fca@umich.edu} 
\affiliation{Michigan Center for Theoretical Physics \\
University of Michigan, Ann Arbor, MI 48109, USA} 

\date{\today}
 
\begin{abstract} 

This paper investigates anthropic bounds on the vacuum energy
$\Lambda$ by considering alternate starting assumptions.  We first
consider the possibility of cosmic observers existing at any random
time (including the future) for constant $\Lambda$, and take into
account the suppression of new structure formation as the universe
approaches its eternal DeSitter (DS) geometry. Structures that
collapse prior to the era of $\Lambda$-domination will lose causal
contact with our Hubble volume within a finite (short) conformal time
$\tau_{\ast}$. Any remnants within our Hubble volume then suffer a
cosmological heat death after the universe becomes DS. The probability
for finding observers by random measurements in the volume bound by
the DeSitter horizon is proportional to the rate of change in the
ratio of the 3-volume $V_{3}(\tau)$ to the 4-volume $V_4(\tau)$, so
that $P \simeq 0$. This vanishing probability of populated DS volumes
is a simple consequence of the information loss problem for eternal DS
spaces resulting from the finite and constant value of its temperature
$T_{DS} \simeq \Lambda^{-1/2}$ and entropy $S = 3/(G\Lambda)$. By
contrast, for geometries with $\Lambda = 0$, structures can condense
and entropy production can continue without bounds at any epoch. The
probability of finding observers in $\Lambda = 0$ geometries is thus
overwhelming higher than in DS spaces. As a result, anthropic
reasoning does not explain the small but nonzero vacuum energy
observed in our universe. We also address the case where observers are
considered only at a specially chosen time -- like the present epoch
-- but relax the allowed values of starting density fluctuations and
hence the redshift of galaxy formation. In this latter case, the
bounds on a $\Lambda$ can be millions of times larger than previous
estimates --- and the observed value. We thus conclude that anthropic
reasoning has limited predictive power.

\end{abstract}


\maketitle

\section{Introduction} 
\label{sec:intro}

This paper investigates the implications of anthropic reasoning on the
selection of the level of vacuum energy $\Lambda$. Our universe is
currently observed to be in an accelerated expansion
\cite{accelerate}, although the nature and origin of the dark energy
that propels the expansion of the universe into an accelerated phase
remains a mystery.  The simplest manifestation of the dark energy
would be a pure cosmological constant $\Lambda$, where $\Lambda
\approx 10^{-122} \, M_p^{4}$ for our universe at the present epoch;
however, vacuum energy that varies with time is also possible.

Both a pure cosmological constant $\Lambda$ and a dynamic mechanism
for dark energy (so that the energy density of the vacuum can be time
dependent) can give rise to similar Hubble expansions at present; as a
result, current observations have difficulty distinguishing between
them.  However, the two models imply crucial differences in the future
evolution of our Hubble volume. A constant vacuum energy density leads
at the classical level \cite{gh} to an eternally DS space in thermal
equilibrium, with an eternally finite and constant bound on the
entropy given by the horizon size, $S_{DS} = 3 / G\Lambda$. In
contrast, a universe with dynamic dark energy has no bound on its
growth of entropy.

Unlike the case of universes with dynamic dark energy components, a
universe with a pure nonzero vacuum energy $\Lambda$ cannot produce
new structures after it thermalizes at the DS temperature $T_{DS}
\simeq \Lambda^{-1/2}$, since the bound given by $S_{DS}$ does not
allow further entropy generation. In addition, any remnant condensates
produced before thermalization at $T_{DS}$ disappear in the DS epoch
since they either lose causal contact with our Hubble volume or suffer
a cosmological heat death within the DS space. The DS volume empties
out within a finite time of about $ H^{-1}$. Any pre-existing
astrophysical structures die out over a somewhat longer time interval.
In the long term, when thermal equilibrium is reached, the temperature
$T_{DS}$ and entropy $S_{DS}$ in this Hubble volume approach and then
maintain a constant value at all times and all points in space; as a
result, no further work can be done and no heat engine can operate. In
the long term future, this space thus suffers from one type of
cosmological heat death of its remnant structures \cite{fred}.

One currently popular attempt at addressing the observed extraordinary
small value of the vacuum energy $\Lambda$ is through anthropic
selection. In this approach, the value of $\Lambda$ is treated as an
environmental parameter that varies in different pocket universes or
causally disconnected regions of the embedding background. The
challenge for any anthropic selection argument is to first identify an
observer and then estimate the probability distribution for measuring
the observed parameters contained in each pocket universe. The hope is
that the most typical value in this distribution for the $\Lambda$
parameter will turn out to be the one detected in our universe,
thereby solving the naturalness issue for the observed vacuum energy
$\Lambda$.  As we discuss below, however, due to the eternal nature of
the thermal equilibrium of DS spaces, the only allowed value for
$\Lambda$ that would avoid the long-term extinction of observers by a
cosmological heat death is the strict value $\Lambda = 0$.

This paper is organized as follows. We first review, in \S II, the
original anthropic argument \cite{weinberg} and several more recent
generalizations.  In \S III, we investigate anthropic arguments by
considering the probability of finding observers randomly at any
cosmological epoch and take into account their cosmological heat
death; these considerations lead to the prediction of $\Lambda =
0$. In S IV, we present a second alternate argument that requires
observers arise before the present epoch, but allows for earlier
structure formation; these considerations allow for huge values of
$\Lambda$. Note that the first of these arguments applies to constant
$\Lambda$, whereas the second allows for dynamic dark energy.
Finally, we conclude in \S V with a summary of our results, a
discussion of the various ways in which anthropic arguments can be
constructed, and their limited predictive power.

\section{Anthropic Reasoning}

\subsection{Weinberg's Proposal}

Well before the acceleration of the universe was discovered in 1998,
in a seminal paper Weinberg \cite{weinberg} found an approximate upper
bound on the vacuum energy, i.e., $\rho_{\Lambda} \le 10^{3}
\rho_{m}^{0}$, where $\rho_{m}^{0}$ is the present matter energy
density; this bound arises because larger values of $\rho_{\Lambda}$
do not allow galaxies to form.  Weinberg argued that since observers
arise from non-equilibrium processes as the result of structure
formation, then cosmological structures must condense (because
observers are present).  However, galaxy formation only takes place
for sufficiently small energy densities of the vacuum, $\rho_{\Lambda}
\le \rho_m$, which is evaluated at the redshift $z_\ast$ when density
perturbations become nonlinear. Relating this condition to present day
values leads to the bound $\rho_{\Lambda} \le (1+z_\ast)^3 \rho_{0}$,
where the subscript refers to present epoch. Since $z_\ast \le 10$,
this argument leads to a bound on $\rho_{\Lambda}$ that is about $2-3$
orders of magnitude larger than the observed value.  Nonetheless, this
bound sufficiently suppresses the value of $\Lambda$ compared to the
``standard'' quantum gravity estimate of $M_p^{4}$, which is about 120
orders of magnitude too large.  This improvement is considered by many
to lend credence to the predictive powers of the anthropic reasoning.

It should be noted that Weinberg's argument selects a special time
$z_{\ast}$ and was based on the criterion for the nucleation of
condensates prior to $\Lambda$ domination. with no reference to the
future evolution of these structures \cite{entropybound}. As such, the
prior knowledge of the parameters $(\rho_m , z_{\ast})$ can be 
considered as fine-tunings of the approach, as we explicitly explore
in \S IV. In addition, the argument remains (nearly) the same if the
cosmological constant $\Lambda$ is replaced by some dynamic dark
energy. Finally, we note that more recent work \cite{martel} updates
the original Weinberg bound in light of the current observations of
$\Lambda$, and reaches similar conclusions.

$\,$ 

\vskip 0.5truein 

\subsection{Anthropic Multiverse and the \\
Probability Distribution of Observers}

The recent observations of the accelerated expansion of the universe,
which may be due to a pure vacuum energy $\Lambda$, in conjunction
with the major theoretical challenges we face in explaining $\Lambda$,
have revived interest in the anthropic reasoning as a possible means
of predicting ``typical'' values for the vacuum energy density. Of
course, the hope is to ``predict'' the observed value. Toward this
end, various proposals for prescribing a probability distribution for
pocket universes and their internal observers have been put forth 
(e.g., see \cite{anthropicprob}). 

For these proposals, the general prescription is to express the
probability $P_j$ for randomly finding an observer in pocket universe
$j$ by the product of the concentration (priors) $p_j$ of $j$-type
bubbles with a selection weight (conditional probability) $f_j$, i.e.,
$P_j = p_j f_j$.  Here, $f_j$ is the weight assigned to the $j^{th}$
bubble, where the weight is proportional to the average number of
observers within the $j^{th}$ bubble so that $f_j \propto N_j$.

One complication in this approach is related to the puzzle of
Boltzmann brains (BB's) \cite{albrecht}. BB's arise from random
quantum fluctuations about equilibrium and outnumber the regular
observers (like us) that arise from non-equilibrium processes traced
back to the nonlinear growth of density perturbation during the matter
domination epoch, perturbations which themselves originate from the
earliest times. If inflation is generically eternal, as believed till
recently, then the number of pocket universes grows without bound.
According to the Everett interpretation, the number of BB's and
regular observers populating the continually produced bubbles will
also grow without bound.  As a result, using the ratio of Boltzmann
brains to that of regular observers to estimate the probability
distribution $P_j$ described above involves a ratio of infinities and
thus becomes undefined without further specification.  Recent attempts
\cite{vilenkin} propose to identify the production of Boltzmann brains
with the bubble abundance factor $p_j$ rather than $f_j$ and identify
observers in $f_j$ only as those arising from non-equilibrium
fluctuations in order to regulate the nucleation rates of BB's and
regular observers.

In the end, however, this anthropic pocket-based approach, like the
original Weinberg argument, is concerned only with the nucleation
rates of the observers (regular observers and BB's), and does not
include the decay rates in estimating the average number of observers
$f_j$ for DS pocket-universes. As we argue below, the inevitable loss
of observers in a DS space-time changes the predictions from this line
of reasoning.

For completeness we note that inflation may not be generically
eternal, as shown recently by \cite{lauraparker, guth}. If this is the
case, the puzzle of bubble population goes away. However, bubble
production by tunneling or through the string theory landscape may
lead to a multiverse picture similar to the one from eternal
inflation.

\subsection{High Complexity Proposal}

Recently, the proposal of \cite{boussoanthropic} uses similar
reasoning to Weinberg's original selection criterion but replaces the
prerequisite of galaxy formation for the nucleation of observers
\cite{weinberg} by hypothesising the requirement of a high entropy content $\Delta
S$ within a causal patch as a sufficient complexity prerequisite to
give rise to observers. The entropy production $\Delta S$ provides the
weighting factor for the population term $f_j$ in the probability
distribution described above using the conjectured principle that
structure is most likely to be found in universes where the values
$\Lambda$ allow the entropy inside a casually connected region to be
maximized. Radiation from stellar burning, reprocessed by dust, is the
major source of entropy in this case. Therefore the amount of entropy
produced depends on the distribution of matter $M$ contained in halos
by time $x=a^3$. The probability $F$ of a given volume being bound at
scale factor $a$ is given by the Press-Schecter (PS) function
\cite{press}, which can be written in the form
\be 
F[\rho_{\Lambda}, M,x, x_e, \mu] \simeq {\rm Erfc} 
\left[ \frac{A(\rho_{\Lambda}/\rhostar)}{G(x) s(\mu)} \right] \, , 
\ee
where $A \simeq 5.6$, $\rhostar = \xi^4 P^3$, $\mu =M \xi^{-2}$, and
$\xi^2=M_e$. In addition, $G(x)$ is the growth function, $P$ is given
by the power spectrum, and $M_e$ is the mass contained within the
volume at the time $x_e$ of matter-radiation equality. This function
has a well-defined peak when its argument is near $10^2 \rhostar$. 
Therefore $\mu$ expresses the mass at time $x$ relative to the horizon
mass scale $\xi^{-2}$ at equality time $x_e$. The horizon mass scale
at the equality time is the crucial physical scale built in the PS
formula.  The lack of the generation of new entropy when the universe
has transited from the matter domination to the $\Lambda$ domination
epoch, is given by the finite conformal time of the causal diamond
$V_3 \rightarrow 0$. Therefore the issue of a finite amount of entropy
spread over an infinite range of physical time, described by the
4-volume $V_4 \simeq H^{-2}$, or equivalently ignoring the influence
of $\Lambda$ on these structures during the DS era, remains as severe
in this proposal as in the previous ones. As an illustration of these
effects, let us mention that photons from stellar burning are
redshifted to infinity when these astrophysical objects are expelled
from the causally connected region. In this sense both arguments
\cite{weinberg, boussoanthropic} are similar. The improvement in the
value of dark energy density \cite{boussoanthropic}, compared to the
value found in \cite{weinberg} comes from requiring that the maximum
entropy estimated with the PS formula be near the maximum volume of
the causal diamond.

Indirectly, this argument contains a fine-tuning of the dark energy
coincidence puzzle. The $\mu$ scale of the PS function introduces the
condition that the matter-dark energy equality time $x_{\Lambda}$
occurs close to the time of matter-radiation equality $x_e$. Requiring
that the coincidence puzzle for $\Lambda$ be satisfied at a special
time, namely when $\rho_{\Lambda} \simeq \rho_m$ near present time,
leads to an improved value for $\Lambda$ compared to the one in
\cite{weinberg}.

\section{Cosmological Heat Death of Anthropic Observers} 
\label{sec:model}

In this section we revisit the limits that anthropic considerations
place on the cosmological constant $\Lambda$. In a $\Lambda$CDM
universe that classically becomes an eternal DeSitter space, the time
during which structure and observers exist is finite and small. To be
precise, we define the existence of observers to be equivalent to the
generation of entropy production. In addition, we focus this
discussion on regular observers (as opposed to BB's) since they (we)
are known to exist.
  
As emphasized above, previous proposals invoking the criterion of
environmental selection have been concerned only with the conditions
that allow for the nucleation of observers, without including the
termination of the observers in the analysis.  This selection has been
done by either using static measures on some spatial hypersurface
(which leads to ambiguous procedures for `time' slicing), or by
assigning to each pocket universe a statistical weight that is further
weighted with the nucleation density of observers in each pocket.  The
weights are generally distributed with respect to the nucleation
density.  Independent of other pathologies (e.g., ambiguities in
defining a global versus local description of measure, ratio of
infinities, the dependence on the choice of observers, and definitions
of the time parameter), these proposals share the property that the
evolution and the decay rate of observers are not taken into account
when constructing a selection criterion.

The motivation for ignoring the influence of the long term fate of
observers --- specifically their decay rate --- on predicting
$\Lambda$ has been based on the fact that once structure forms than it
decouples from the Hubble flow. On this basis, it was expected that
$\Lambda$ would have no influence on the evolution of the stars or
observers after the universe reaches its DS phase. 

Ignoring the influence of $\Lambda$ on the existence of observers is
not correct for the DS spaces due to the eternally constant and finite
values of its temperature $T_{DS}$ and its entropy $S_{DS}$: The
number of observers produced before the DS epoch decreases
exponentially due to the superluminal expansion as they get expelled
from causal contact, or equivalently since every co-moving observer
sees an effective shrinking horizon. 

As is well known \cite{gh}, in about a Hubble time the physical radius
of our universe approaches the constant value $H^{-1} \simeq
\Lambda^{-1/2}$ and our universe approaches thermal equilibrium from
that time onward.  As a result, the universe eventually maintains a
constant temperature $T_{DS} \simeq H^{-1}$ at all points and a
constant finite entropy $S_{DS}\simeq 3/\Lambda$ for all (sufficiently
late) times. After such an equilibrium is reached, the universe loses
its ability to do physical work, and no heat engine can operate
because the energy available is finite and is distributed over an
infinite time interval. The universe and any observers it contains
thus suffer from one kind of {\it cosmological heat death}.  This
behavior is a manifestation of the well-known information loss puzzle
for DS states, since the finite constant value of the temperature and
entropy leading to this cosmological heat death of observers are a
consequence of the fact that the DeSitter horizon does not evaporate
away (unlike black holes).  If the horizon could evaporate away, then
the temperature and the entropy of the DS space would not maintain
their constant value, and this type of heat death would not occur.

Cosmological heat death \cite{fred} is the modern counterpart to
classical heat death \cite{heat}, which, roughly speaking, would take
place if the universe reached a thermodynamic equilibrium. In such a
state, the universe would have the same temperature at all points in
space, so that no heat engine could operate and no work could be
done. The expansion of the universe changes the concept of heat death
as follows: For universes that are not vacuum dominated, the
background temperature of the universe is continually decreasing, so
that the universe can never reach the constant temperature state of
classical heat death.  Nonetheless, such a universe can become purely
adiabatic, so that no more work can be done within it, and can thus
suffer one type of cosmological heat death \cite{fred}; however, since
there are no bounds on the entropy, observers can be continually
produced via the flow of density perturbations accross the horizon,
i.e., the universe need not become adiabatic.  A DS universe, in
contrast, can eventually reach a constant temperature state (at the DS
temperature), and can thus suffer a cosmological heat death of another
type. This second kind of cosmological heat death is the one relevant
for investigating anthropic arguments for $\Lambda$ that consider 
observers at arbitrary cosmic epochs (as discussed here).

As argued above, any remnant observers in a DS universe end up in a
cosmological heat death (of the second kind). Due to the constant
finite entropy $S_{DS}$, no new entropy and hence no new observers can
be generated. For these reasons, the universe in its DS epoch has no
production channels for observers or entropy over an infinite time
interval, while all the pre-DS epoch observers disappear within a
short finite time. On the largest spatial scales, the cosmological
heat bath shuts down entropy production on a time scale $\delta \tau
\simeq \Lambda^{-2/3}$. On smaller scales, extant astrophysical
objects (e.g., stars and stellar remnants) continue to generate energy
and entropy until their fuel sources are exhausted. Although these
time scales are somewhat longer than that of the cosmological heat
bath \cite{fred}, the lifetimes of these bodies are nonetheless
finite.

It is straightforward to quantify this statement concerning the
influence of $\Lambda$ on the number of observers. Let us denote by
${\cal N} = n a^3$ the number of observers in the Hubble volume, where
$n$ is the number density and $a(\tau)$ is the scale factor, with
$\tau = \int dt/a $ the conformal time and $t$ physical time. The
(co-moving) 3-volume is given by $V_3 =(4\pi/3) [\tau_\ast - \tau]^3$,
where $\tau_\ast$ is the future asymptotic value of the conformal time
in our causal domain.  The 3-volume $V_3$ thus approaches zero in the
limit $t \to \infty$ ($\tau \to \tau_\ast$).  This behavior represents
the `shrinking of the horizon' in co-moving coordinates. On the other
hand, the (co-moving) 4-volume $V_4 =\int V_3 dt \propto H^{-2}$ has a
finite value determined by the physical radius of the Hubble volume
$H^{-1}$. The total number of observers in the co-moving volume is
roughly proportional to the volume ${\cal N} \simeq V_3 $, so that it
decreases with the `shrinking of the horizon' of $V_3$. The number of
observers thus gets smaller according to 
\be 
{d{\cal N} \over dt} \simeq {dV_3 \over dt} = 3 {\cal N} 
{1 \over \tau - \tau_\ast} {d \tau \over dt} + \Gamma {\cal N} \, . 
\ee 
Next we can define the parameter $x$ through 
\be 
x \equiv - {1 \over \tau - \tau_\ast} {d \tau \over dt} 
 \propto - \frac{\dot {V}_3}{3 \dot{V}_4} \, , 
\ee
and find that the balance equation is
\be
{d V_3 \over dt} = - 3 x V_{3} \, . 
\ee 
Note that $x > 0$ since $d\tau/dt$ is negative, and it varies between
$H/2$ in the past and $0$ at future infinity. We have added a particle
production ($\Gamma>0$) or annihilation ($\Gamma<0$) term for the most
general case. If continued generation of entropy production were
possible, then the above equation for the number rate would be
modified by a source term $S(t)$ so that $d{\cal N}/dt \simeq dV_3 /dt
+ S(t)$. But since the continued generation of entropy over the whole
infinite time interval is strictly zero for DS spaces after reaching
thermal equilibrium, we have $S(t)=0$. If $\Lambda$ is not strictly a
cosmological constant, but rather a dynamic dark energy component,
then quite possibly $S(t) \ne 0$ would contribute to the generation of
observers. Ignoring $S(t)$ we can equivalently write this differential
equation, as a balance equation in terms of the number density of
observers $dn/dt +3Hn = -3x n + \Gamma n$. Due to the cosmological
heat death and thermal equilibrium in the Hubble volume then $\Gamma
\equiv 0$. Solving the differential equation we find that ${\cal N} =
na^3 = {\cal N}_0 e^{-3xt}$ which reflects nothing more that the
dilution of all objects expelled from our Hubble volume due to the
accelerated expansion.

In short, we have shown that the number of observers $\langle {\cal
  N}_j \rangle = f_j$, crucial for the anthropic predictions for the
distribution of the values of $\Lambda$ in $j$ bubbles, approaches
zero in any space with a nonzero vacuum energy $\Lambda$ over an
infinite time interval. Observers exist for a brief time interval
around the $\Lambda$ domination era. By applying the Copernican Time
Principle for the past, present, and future of our universe\cite{fred}, 
which states that there is nothing special about our present epoch for
picking this time to make a measurement, then making a measurement
randomly in space and time in a DeSitter universe $j$ should yield an
identically zero number of observers $f_j = \langle {\cal N}_j \rangle
= 0$. Constraining ourselves to the selection of a special timeslice
when observers are known to exist, instead of a random one for making
a measurement, then counts as fine-tuning rather then a prediction of
the theory. 

Although in this note we are not concerned with the viability of eternal inflation, it should be noted that our argument for the cosmic heat death of observers can be extended to the case of the eternal inflation for the following two reasons: i) the anthropic approach to the probability of the eternal inflation is given by the product of the number of bubbles with the number of observers in them, as reviewed above. Although our argument has nothing new to say about the number of bubbles produced, it does yield a vanishing number of observers in each bubble, thereby making the anthropic probability given by the product equal to zero; ii) the global boundary condition we consider for this approach considers a timeslice, taken at time infinity. Since the parameter of time is conventionally taken to be in common for all the bubbles produced by eternal inflation then the timeslice in the far future would cut through all the bubbles.Since observers come into existence for a very brief time during the history of the bubble, then in this hypersurface at future infinity, the number of bubbles empty of observers will be overwhelmingly larger then the bubbles where observers are accidentally found.
 
We thus conclude that anthropic considerations (without additional
assumptions, e.g., a preferred measurement time) are not consistent
with the observed value of $\Lambda$. As a result, such arguments
cannot be invoked as a means of predicting values of $\Lambda$
different from zero since the observers, which are at the heart of the
anthropic principle, do not exist in the cold empty equilibrium DS
spaces.

\section{An Alternate Argument Allowing for Large Vacuum Energy Density} 

As illustrated by the previous section, the basic point of this paper
is that anthropic ``predictions'' depend on what is assumed as a
starting point (the ``priors'' in the language of statistics). The
above argument (\S III) allows for observers to be considered at any
time, and then finds that the existence of observers essentially
requires a zero value for the cosmological constant $\Lambda$.  In
this section for completeness we review the case when $\Lambda$ is not strictly a cosmological constant but a dynamic energy component, in which case the entropy bounds of DS spaces would not be applicable. Here we require that the universe produce observers before
the present time (as in \cite{weinberg}), but allow the redshift of
structure formation to change, and find that the value of $\Lambda$
can be millions of times larger than the observed value. A dynamic  $\Lambda$ case has also been discussed in \cite{aguirre}) with similar conclusions to the ones here.

We note that the higher redshifts for structure formation considered here
would require correspondingly larger density fluctuations in the early
universe and hence larger temperature fluctuations in the cosmic
background radiation.  These alternate universe would not share all of
the properties of our universe --- only the requirement that observers
can be produced.

For completeness we also note that the previous argument of \S III
applies specifically to the case in which $\Lambda$ is strictly
constant. The situation may change if the dark energy density is time
dependent, since there are no bounds on the entropy increase in this
case, and hence the generation of observers may not be forbidden at
randomly selected times for making a measurement. However, the
argument constructed in this section applies to both the case of a
cosmological constant and a dynamic dark energy component.

Specifically, the arguments of \cite{weinberg} start with the idea
that large values of the vacuum energy $\rho_V$ (or, equivalently,
$\Lambda = 8 \pi G \rho_V$) suppress galaxy formation. As a result, 
in order to form nonlinear structures, the vacuum energy density 
must obey the bound
\be
\rho_V < {500 \over 729} \rhowiggle \, , 
\label{eq:vacbound} 
\ee 
where $\rhowiggle$ provides a measure of the starting size 
of the density perturbations $\Delta \rho$, i.e., 
\be 
\rhowiggle \equiv \lim_{t \to 0}  
\left\{ { [\Delta \rho (t)]^3 \over \rho^2 (t) } 
\right\} \, . 
\label{eq:wigdef} 
\ee 
After further algebra \cite{weinberg}, one finds that 
the formation of structure (galaxies or quasars) before 
a redshift $z_\ast$ provides a lower bound on the maximum 
value of the parameter $\rhowiggle$ defined above: 
\be
\rhowiggle\Big|_{max} \ge {729 \over 1500} \pi^2 \rho_0 
(1 + z_\ast)^3 \, , 
\label{eq:wigbound} 
\ee
where $\rho_0$ is the present-day density of the universe. 

The original argument invoked the fact that our universe is observed
to have produced quasars before a redshift of $z_\ast \sim 4.5$. Using
this value in equation (\ref{eq:wigbound}), in conjunction with
equation (\ref{eq:vacbound}), implies that the anthropic upper bound
on the vacuum energy density is {\it at least as large as} $\sim550
\rho_0$.  However, this version of the argument relies on the choice
of $z_\ast \sim 4.5$ in order to find a numerical value (see also
\cite{aguirre}). In the vast ensemble of possible universes, other
regions of space-time could form quasars earlier. Thus, one could
consider larger values of the formation redshift. If we use $z_\ast
\sim 100$, for example, then the bound gets weaker, so that we would
only conclude that $\rho_V$ is less than $\sim$5,000,000 $\rho_0$, a
considerably weaker result.

Relaxing the redshift requirement for quasars thus allows for a much
larger vacuum energy density than the original anthropic argument of
\cite{weinberg}, and a much larger value than we observe.  One can
even take this argument to its limit, and consider universes in which
structure forms right after decoupling, which correspond to a redshift
of $z_\ast \sim 1000$, so that the resulting bound becomes $\rho_V <
10^9 \rho_0$.

Note that this argument applies equally well to approaches that rely
on the PS formalism. The PS formula is an error function that peaks
around $10^2 \rho_{\ast}$, which is the matter energy density at the
time of matter-radiation equality. The special scale built into this
formula is thus the redshift when the matter radiation equality
occurs.  We can easily envision, in the vast multitude space-time
regions in the multiverse, universes with a different matter
inventories, so that the matter-radiation equality occurs at very
different redshifts from our universe. In this case, the probability
described by the PS formula peaks at different values since the scale
$\rho_{\ast}$ varies. As a result, the predicted value of $\Lambda$
(as function of $\rho_{\ast}$) peaks at different values than in the
case of our universe.

To summarize, we have argued that anthropic arguments with one set of
assumptions (\S III, where observers are considered at all cosmic
times) implies that $\Lambda$ = 0, whereas another set of assumptions
(here, where cosmic structure can form earlier) predicts, or at least
allows, $\Lambda$ to be a billion times larger than its observed
value. The predictions of this class of anthropic arguments thus
depend sensitively on the starting assumptions. This sensitivity,
coupled with the lack of a way to specify {\it a priori} what the
``correct'' assumptions should be, severely limits the predictive
power of anthropic reasoning. As a result, the anthropic approach
becomes an unfalsifiable theory.

\section{Conclusions}

This paper has investigated the viability of anthropic reasoning for
predicting the value $\Lambda$ of the vacuum energy density of the
universe.  This discussion differs from previous considerations by
incorporating the boundary conditions of the observers in eternally DS
spaces, in particular, the fact that a finite amount of entropy is
spread over an infinite time interval.  Since entropy production is
necessary for the continued existence of observers, the inability of
universes with $\Lambda \ne 0 $ to continue entropy production
ultimately leads to the lack of observers. Specifically, we showed
that the observers in any eternal DeSitter universe ($\Lambda \ne 0$)
suffer a cosmological heat death in the long term future.  This
finding, together with the fact that no continued entropy production
is allowed, implies that the number of observers found by any given
random measurement would be zero. Because of this lack of observers in
DS spaces, anthropic approaches imply values of the vacuum energy
density $\Lambda$ that are strictly zero with an overwhelmingly larger
probability, proportional to the ratio of the 3-volume over the
4-volume, a prediction that is not consistent with current
observations of our universe.


In contrast to previous assumptions, $\Lambda$ has a significant
influence on the existence of observers, even though nonlinear
structures (halos) separate from the Hubble flow.  Because the
accelerating universe effectively separates (present-day) dark matter
halos into island universes in the future \cite{island}, each such
bound structure has finite energy and entropy resources, and hence a
decaying number of observers. Any remaining observers also suffer a
cosmological heat death. Thus, because of the vacuum energy, random
measurements performed at any randomly picked time would typically
find the number of observers in spaces with nonzero vacuum energy to
be zero. In this sense, anthropic considerations do not predict the
observed value of $\Lambda \ne 0$. For completeness, we note that
these considerations do not apply for open universes; unlike the
spatially flat DS space with a fixed physical radius $H^{-1}$, open
universes lack a spatial bound and have no limit for their entropy and
temperature. 

This severe measure problem lies at the heart of the information loss
puzzle for DeSitter universes. Taking into account the eternal nature
of DeSitter space \footnote{ At least at the classical level \cite{gh}
DS space is eternal with constant and finite temperature and entropy.
This solution may change when quantum effects and back reactions on
the DS geometry are taken into account \cite{arrowlaura, don}. In the
latter case, DS space may decay in astronomical scales due to
instabilities from quantum effects.}, in combination with the fact
that DS space soon empties out of observers and all its matter
content, leads to the conclusion that most observers nucleated before
vacuum energy domination are soon after lost from causal contact; the
ones that survive into the DS era then suffer a cosmological heat
death.  Furthermore, no net production of observers in the eternal DS
epoch is allowed since their nucleation would contribute an entropy
channel to the DS entropy. At all times the observers decay rate is
at least as large as the nucleation rate in DS spaces. For example, in
our universe protons decay with a time scale in the range $10^{33} -
10^{45}$ yr, where the lower limit comes from current experimental
bounds \cite{superk} and the upper limit is theoretical \cite{fred}.
Similarly, black holes evaporate on a time scale roughly given by
$\tau = 10^{65}$ yr $(M_{bh}/M_\odot)^3$, so that even the largest
black holes are gone after $\sim 10^{92}$ yr; these objects provide
the last important contributions to the background non-equilibrium
processes for astrophysical remnants inside the causally connected
region \cite{fred}. From then to infinity the Hubble volume attains
thermal equilibrium with the same temperature $T_{DS}$ at all points
and time.

Using a Copernican Principle approach, that is, randomly making
measurements for observers in DS spaces, it then follows from above
considerations that the probability to randomly find an observers at
any time has a measure zero.  This result is independent of how the
time parameter is defined. This argument is, in essence, a different
way of stating the well known problem of information loss, but it is
applied it here to anthropic considerations for predicting the typical
values for $\Lambda$. As outlined above, anthropic selection thus
requires $\Lambda$ to be strictly zero, in conflict with the observed
value for our universe.


The difference between previous anthropic predictions and those of
this paper can be summarized as follows: Previous arguments assume
that the only observers that should be considered are those in the
present day universe, whereas this paper adopts a more global view,
with observers being allowed to exist at any epoch. If we limit the
possibilities from the onset to the case of finding observers in a
young universe, $t < 14$ Gyr, then a small nonzero value of $\Lambda$
does not hurt the odds; however, if we consider the question of
observers in the universe at any time, then any small nonzero value of
$\Lambda$ is devastating. So the question of whether or not anthropic
arguments imply, or even allow, nonzero values of $\Lambda$ depends on
whether the present cosmological epoch is special or not. In other
words, the predictions of anthropic arguments depend on whether or not
one adopts the Copernican Time Principle. The predictions of anthropic
arguments can be radically different for varying initial assumptions
and thus their predictive power is severely limited.

\medskip 

Acknowledgment: We would like to thank Lee Smolin and Dragan Huterer
for useful discussions, and the organizers of the 2007 fq(x)
conference, Max Tegmark and Anthony Aguirre, for helping to facilitate
this collaboration. FCA is supported by the Michigan Center for
Theoretical Physics and by the Foundational Questions Institute
through grant RFP1-06-1. LM-H is supported in part by DOE grant
DE-FG02-06ER1418 and NSF grant PHY-0553312.


\end{document}